# MindBigData 2022: A Large Dataset of Brain Signals

David Vivancos and Félix Cuesta

Email: *vivancos@vivancos.com*  *felix@felixcuesta.com*

**Abstract**

*Understanding our brain is one of the most daunting tasks, one we cannot expect to complete without the use of technology. MindBigData [1] aims to provide a comprehensive and updated dataset of brain signals related to a diverse set of human activities so it can inspire the use of machine learning algorithms as a benchmark of "decoding" performance from raw brain activities into its corresponding (labels) mental (or physical) tasks. Using commercial of the self, EEG [2] devices or custom ones built by us to explore the limits of the technology. We describe the data collection procedures for each of the sub datasets and with every headset used to capture them. Also, we report possible applications in the field of Brain Computer Interfaces or BCI that could impact the life of billions, in almost every sector like healthcare game changing use cases, industry or entertainment to name a few, at the end why not directly using our brains to "disintermediate" senses, as the final HCI (Human-Computer Interaction) device? simply what we call the journey from Type to Touch to Talk to Think [3].*

## 1.- Introduction

The brain is said to be our final inner frontier, where all our experiences are built and what generates our conscience, but the tools we have so far to try to understand it, are to say the least very rudimentary due to several reasons, to name the main two:

First the complexity our brain with close to 100 billion neurons [4] and trillions of connections (synapses) using electrical and chemical processes, a true web of intricate relationships that we are still in its early days of exploration, without a definitive model or models that could describe even in some basic detail its inner workings.

Second due to the tremendous limits of our current technology to "read" (or write) brain activities in real time, in a lab setting or "in the wild", at the time of the writing of this document our current technology is several orders of magnitude below human raw capabilities in terms of spatial resolution and limited in terms of temporal resolution.

In this paper we introduce an EEG group of datasets called "MindBigData", a large-scale ontology of brain signals. As a continuous approach to capture labeled brain signals for almost a decade, started in 2014.

There are 3 main "MindBigData" databases:

1.- "The MNIST [5] of Brain Digits" for EEG signals with several headsets captured while looking at "font" based digits shown in a screen from 0 to 9.

2.- "The ImageNet [6] of the Brain" for EEG signals captured while looking at ImageNet images.

3.- "The Visual MNIST of Brain Digits" for EEG signals captured while looking at the original MNIST handwritten digits (real pixels) shown in a screen also from 0 to 9.

The paper includes a description off all the datasets in section 2, details on how it is built in section 3, possible applications in section 4, future prospects in section 5 and a selection of relevant citations by third party papers in section 6.

## 2.- Properties of MindBigData

Each of the MindBigData Datasets is composed of one or several datafiles for every EEG headset used to capture the brain signals, the files hold the data for each of the channels available, dependent on the device used to capture, from 1 channel in the simplest device to 64 channels in the highest density one, also the labels are included to correctly tag the brain activity related to each event, recording time, brain locations, sampling rate and other details are explored in the following subchapters, with specific information for each of them.

With the release of this paper an additional splitting between train & test datasets has been included for unified benchmarking purposes, detailed in chapter 2.4.

## 2.1- MNIST of Brain Digits

This is the first MindBigData dataset created since 2014 and the one cited in more papers or subsequent related research by third party researchers.

It consists of 4 datasets each captured with a different commercial EEG headset, with varying channel density, from 1 Channel to 14 EEG Channels placed in different areas of the brain using the 10/20 [7] naming convention, in Frontal, Parietal, Temporal and Occipital brain regions, each showed in the following FMRI [8] image from the real brain of the experiment subject, with labels in different colors:

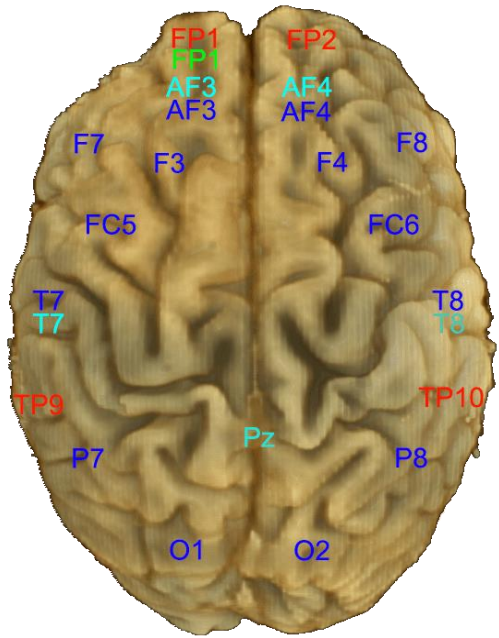

NeuroSky [9] MindWave 1 (2011) with a single EEG Channel (FP1 shown in green in the picture above), and 67,635 brain signals of 2 seconds each captured, at a theoretical sampling rate of about 512 samples per second or 512Hz, last dataset version is v1.0.

Interaxon [10] Muse 1 (2014) with 4 EEG Channels (FP1, FP2, TP9 & TP10 shown in red in the picture above) and 163,250 brain signals of 2 seconds each captured, at a theoretical sampling rate of about 220 samples per second or 220Hz, last dataset version is v1.01.

Emotiv [11] Insight 1 (2015) with 5 EEG Channels (AF3, AF4, T7, T8 & Pz shown in light blue in the picture above) and 65,250 brain signals of 2 seconds each captured, at a theoretical sampling rate of about 128 samples per second or 128Hz, last dataset version is v1.06.

Emotiv EPOC 1 (2009) with 14 EEG Channels (AF3, AF4, F7, F8, F3, F4, FC5, FC6, T7, T8, P7, P8, O1 & O2 show in dark blue in the picture above) and 910,476 brain signals of 2 seconds each captured, at a theoretical sampling rate of about 128 samples per second or 128Hz, last dataset version is v1.01.

Sampling rate for some of the devices is usually not 100% constant, it can be more, or it can be less, so one option is to resample, interpolating for example to match the ideal sampling rate.

The signal measures variation in voltages from brain neural activities and it's provided in "raw" format as it was captured from each of the devices, so value ranges are relative to the device, some are integers some floats.

The brain signals were captured while the subject was for 2 seconds seeing a single digit (showed on a 65" TV screen in a white font over a full black background) from 0 to 9 like:

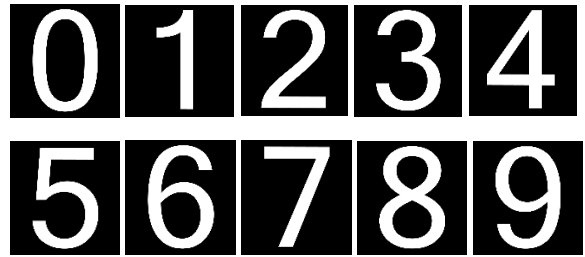

The appearance of the digits was random with a black screen in between them.

59,339 signals were also captured while doing other mental activities not related to digits, without any limitation on movements, blinks, or other actions, these were labeled as "-1". This is present for all the headsets but Emotiv Insight.

The structure of the files is the same for all, with no headers in the original files, the format is simple text, and the fields are separated by a "TAB" character.

There are 7 columns to describe the data:

The 1st column is "**id**", numeric integer for reference purposes only.

The 2nd column is "**event id**", numeric integer to distinguish a singular event but captured at different brain locations, used only by multichannel devices, note that to differentiate the channels the 4th column is used.

The 3rd column is "**device**", a 2 characters string to identify the device used to capture the signals, "MW" for MindWave, "MU" for Interaxon Muse & "IN" for Emotiv Insight, "EP" for Emotiv EPOC.

The 4th column is "**channel**", a variable size character string to identify the 10/20 brain location of the signal, with possible values:

For MindWave: "FP1".

For Interaxon Muse: "TP9","FP1","FP2" or "TP10".

For Emotiv EPOC: "AF3, "F7", "F3", "FC5", "T7", "P7", "O1", "O2", "P8", "T8", "FC6", "F4", "F8" or "AF4".

For Emotiv Insight: "AF3","AF4","T7","T8" or "PZ".

The 5th column is "**code**", numeric integer to distinguish the digit been seen & thought, with possible values 0,1,2,3,4,5,6,7,8,9 or -1 for random captured signals not related to any of the digits.

The 6th column is "**size**", numeric integer to identify the number of samples captured for this event, in theory should be 2 (seconds) by the sampling rate, but for some devices sampling rate is variable, and this size is not a fix number.

The 7th column is "**data**", a set of "n" numeric integers or floats (if so, using a dot "." as radix point to separate the integer part from the fractional part), being "n" a variable value indicated in the previous column "size" and to separate the numbers the character coma "," is used as the delimiter inside this column.

There are 395,072,896 Data Points, in all the files of this dataset "MNIST" of Brain Digits.

These datasets can be downloaded at:

http://mindbigdata.com/opendb/index.html

December 2022 curated versions of this datasets at Hugging Face, are explored in chapter 2.4.

## 2.2- Imagenet of the Brain

Started in 2018, current v1.04 contains 70,060 brain signals of 3 seconds each captured with Emotiv Insight 5 channels EEG (AF3, AF4, T7, T8 & Pz shown in light blue in the picture at chapter 2.1), while the subject was seeing a random image from the Imagenet ILSVRC2013 [12] training dataset, covering in total 14,012 images.

There are two datasets one with only the raw EEG waves and another including additionally a spectrogram (only for 10,032 of the Images generated using the brain signals captured) and included as an extra image-based dataset.

### 2.2.1- EEG Data Files

The structure of the files is as follows:

There is one csv [13] file for each EEG set of signals related to a single Imagenet image.

The naming convention if we take for example the file:

"MindBigData_Imagenet_Insight_n09835506_15262_1_20.csv"

**MindBigData_Imagenet_Insight_** : relates to the EEG headset used Insight atm only.

**n09835506** : relates to category of the image from the "synsent of ILSVRC2013" in this example n09835506 is "ballplayer, baseball player".

Additional Note: It was added a "WordReport-v1.04.txt" file too in the zip file with 3 files per row TAB separated with: the category names, the eeg image recorded count and the synsent ID)

**15262** : relates to the exact image from the above category, all the images are from the ILSVRC2013_train dataset, downloaded from the Kaggle Website. This image for example is n09835506_15262.JPEG at the folder:

ILSVRC2013_train\n09835506\

**_1_** : relates to the number of EEG sessions recorded for this image, usually there will be only 1 but it is possible to have several brain recordings for the same image, second will be 2 and so on.

**_20** : relates to a global session number where the EEG signal for this image was recorded, to avoid long recording times only 5 images are shown in each session with 3 seconds of visualization and 3 seconds of black screen between them.

Inside each of the csv files there are 5 lines of plain text one for each EEG channel recorded, ending with a new line escape character.

Each line starts with the Channel name from the Emotiv Insight Headset as a text "AF3, "AF4", "T7", "T8" or "Pz".

And then separated by (,) for this headset data rate is at 128Hz so there should be around 384 (128 x 3 secs) decimal values like "4304.61538461538" for each channel note the dot is used for the decimal point. Since the headset could provide more or less samples, 384 is just a reference.

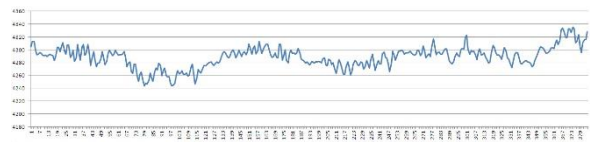

The above plot is generated to show the data from a single EEG channel over 3 seconds.

All the 5 EEG channels follow the same pattern, and the "time" coordinate of the time series is shared between the 5 channels so after the channel labels the next "column" of numbers is the first time-step and so on.

There are 26,903,040 Data Points, in all the files of this dataset Imagenet of the Brain

### 2.2.2- EEG Spectrogram Data Files

For 10,032 of the Imagenet images or dataset version 1.0, using the raw EEG data, a spectrogram [14] was generated with a custom color encoding having the following structure:

The spectrogram is created using only 3 of the 5 EEG channels available, creating PNG [15] RGB [16] Files, AF3 channel is used as RED, AF4 as GREEN, and Pz as BLUE, this is a custom choice and others will be interesting to explore.

To generate the spectrogram for each raw EEG wave, the first 64 samples (1/2 seconds) are discarded to avoid possible communication lags, and then 128 samples are used to generate an FFT [17], and so on for each time step.

Each time step is taken with a move ahead of 2 samples, until we reach 256 samples (2 seconds) so in total we cover 64 overlapped timesteps too. Probably the overlap is too big you may want to try smaller ones in your pre-processing pipeline.

Notice that for this sample scenario we are using the raw wave, but it is advisable to use some filters previously as existing EEG literature suggest [18].

At the end we have 64 Frequencies (0 to 63hz) values (from the FFTs of 128 samples) for each of the overlapped 64 timesteps.

Notice also that the frequency range here is not limited and include some frequencies beyond what is expected for a "typical" brain signal. probably EMG [19] or other signal artifacts.

Once we have the frequency (magnitude) values, they are scaled and coded into color values (0-255) using EEG AF3 channel as RED, AF4 as GREEN and Pz as BLUE.

To reduce the effects of outliers, the frequency value distribution is proportionally mapped into the 0-255 value range for each color channel.

Here is a sample spectrogram amplified 10 times (the ones included in the zip are only 64x64 pixels)

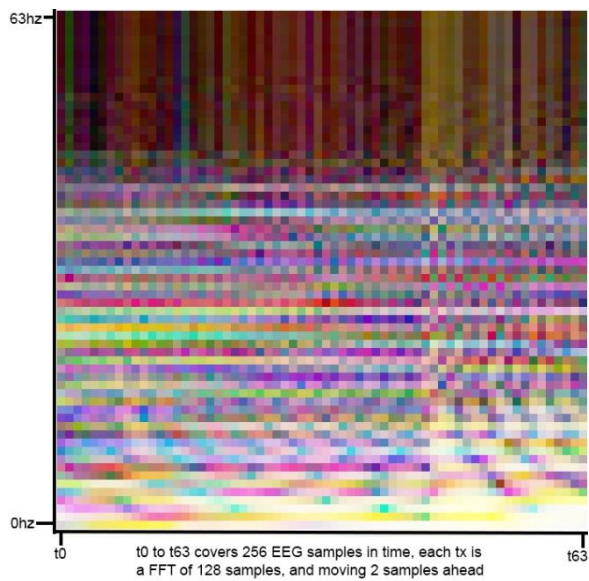

The name of the files will be the same as the csv but ending in ".png" and you can find them in the folder MindBigData-Imagenet-v1.0-Imgs

Beware that a few of the PNG files maybe filled mostly with a single color reflecting a probable unexpected capture flaw for the EEG device, and probably worth discarding.

No data cleaning or was performed on the original Datasets,

These datasets can be downloaded at:

http://mindbigdata.com/opendb/imagenet.html

December 2022 curated versions of this datasets at Hugging Face, are explored in chapter 2.4.

## 2.3- Visual MNIST of Brain Digits

In 2021 a new dataset was introduced, again with the MNIST digits but now instead of using a regular "font" to be shown in the screen as stimulus with the numbers 0 to 9, the real pixels of numbers from the original MNIST dataset of handwritten digits where used, like these ones:

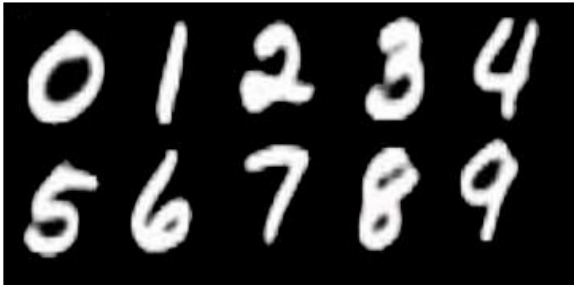

The experiment setup was like the one described in chapter 2.1 but in this dataset the data labeled as -1 was recorded also in a still position, trying to prevent any muscular movement too, but without the visual stimulus of seeing the digit.

The visual MNIST was captured with several EEG headsets:

### 2.3.1- With the Muse2 headset

This dataset, started in 2021 and latest version v.017, it contains 72,000 brain signals of 2 seconds each captured with Interaxon Muse2 headset with 4 EEG Channels (AF7, AF8, TP9 & TP10) and 3 PPG [20] Channels (ambient, Infrared and red), 3 Accelerometer [21] Channels and(X,Y,Z) and 3 Gyroscope [22] Channels (X,Y,Z), that were included to add extra sensors info.

There are 3 data files, one with 18,000 unique digits captured and all the sensor data, and 2 removing noisy captures one with 11,387 digits but only including the TP9 & TP10 EEG channels called "cut2" and another with 1,186 digits only including the TP9, AF7 & TP10 EEG channels called "cut3". In all the datafiles the other corresponding non-EEG sensor data was included.

The data files are stored in a very simple text format without headers and csv like, (,) comma separated with the following fields:

[**dataset**]: a simple text pointing to the original Yann LeCun MNIST source type, can be "TRAIN" or "TEST", related to the 60,000 original train digits and 10,000 test digits.

[**origin**] 1 integer, used to reference the Yann LeCun MNIST location of the original digits in the source data files from 0 to 59,999 for train 0 to 9,999 for test or -1 to indicate black signal (meaning a black image not coming from the original MNIST datasets).

[**digit_event**]: 1 integer with the original MNIST label of the image from 0 to 9 or -1 to indicate black signal (no digit shown).

[**original_png**]: 784 integers (comma separated), with the original pixel intensities from the Yann LeCun MNIST from the source PNG files shown, each pixel can have a value from 0 to 255, (for black signal all will be 0s) 784 comes from (28x28) since it is single channel square image, flattened.

[**timestamp**]: 1 Unix Like timestamp [23] for initial time of recording of the signals for this digit capture.

[**EEGdata**]:

512 floating point (comma separated) EEG - TP9 channel raw signal (2secs at 256hz), followed by

512 floating point (comma separated) EEG - AF7 channel raw signal (2secs at 256hz), followed by

512 floating point (comma separated) EEG - AF8 channel raw signal (2secs at 256hz), followed by

512 floating point (comma separated) EEG - TP10 channel raw signal (2secs at 256hz)

Note: For Muse2 Cut2 reduced data files, only TP9 & TP10 were included and for Muse2 Cut3 only TP9, AF7 & TP10.

[**PPGdata**]:

512 floating point (comma separated) PPG1 ambient channel raw signal (2secs at 256hz), followed by

512 floating point (comma separated) PPG2 infrared channel raw signal (2secs at 256hz), followed by

512 floating point (comma separated) PPG3 red channel raw signal (2secs at 256hz)

[**Accdata**]:

512 floating point (comma separated) Accelerometer X channel raw signal (2secs at 256hz), followed by

512 floating point (comma separated) Accelerometer Y channel raw signal (2secs at 256hz), followed by

512 floating point (comma separated) Accelerometer Z channel raw signal (2secs at 256hz)

[**Gyrodata**]:

512 floating point (comma separated) Gyroscope X channel raw signal (2secs at 256hz), followed by

512 floating point (comma separated) Gyroscope Y channel raw signal (2secs at 256hz), followed by

512 floating point (comma separated) Gyroscope Z channel raw signal (2secs at 256hz)

For this dataset to unify signal length all samples were interpolated to match 512 samples or 2 seconds at 256hz the same used for the EEG but the sources where 64hz for PPG and 52hz for Accelerometer & Gyroscope.

Including the interpolation there are 133,992,000 Data Points, in the main file of this dataset Visual MNIST of Brain Digits using the Muse2 headset.

### 2.3.2- With the custom built 64 channels cap

This dataset started in 2022, latest version v0.016, contains 148,736 brain signals of 2 seconds each recorded with a custom-built EEG cap connected to 2 Contec [24] KT88-3200 amplifiers, and captured with a custom developed software to integrate and synchronize the couple 32 channels amplifiers together to obtain the final 64 EEG Channels, with a sampling rate of 200hz.

The 64 EEG Channels used are: FP1, FPz, FP2, AF3, AFz, AF4, F7, F5, F3, F1, Fz, F2, F4, F6, F8, FT7, FC5, FC3, FC1, FCz, FC2, FC4, FC6, FT8, T7, C5, C3, C1, Cz, CCPz, C2, C4, C6, T8, TP7, CP5, CP3, CP1, CPz, CP2, CP4, CP6, TP8, P7, P5, P3, P1, Pz, P2, P4, P6, P8, PO7, PO5, PO3, POz, PO4, PO6, PO8, CB1, O1, Oz, O2 and CB2.

Like in the 10/20 head distribution bellow

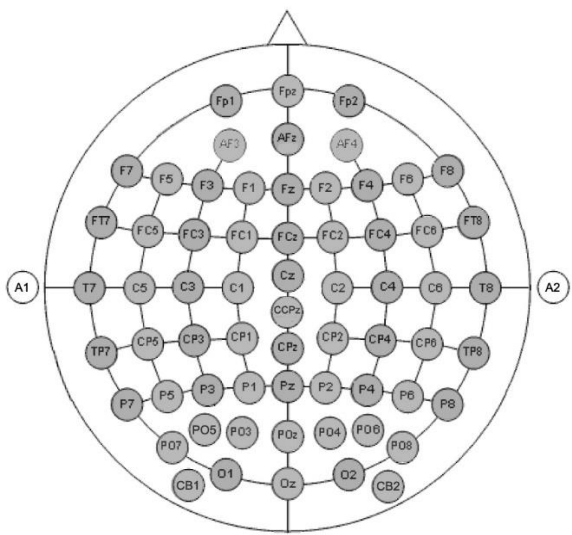

A1 Earlobe clip channel was used as reference [25] for the left side of the brain and A2 Earlobe clip channel was used for reference for the right side of the brain, the center channels CCPz, CPz, Pz, POz and Oz were referenced to A1 and FPz, AFz, Fz, FCz and Cz to A2.

There is 1 single data file with the resulting 1,162 unique digits captured and 1,162 while showing a black screen in between the digits and labeled -1. Using 1 single row per capture totaling 2,324 rows.

The data file is stored in a very simple text format without headers and csv like, with 26,388 (,) comma separated values per row with the following fields:

[**dataset**]: a simple text pointing to the original Yann LeCun MNIST source type, can be "TRAIN" or "TEST", related to the 60,000 original train digits and 10,000 test digits.

[**origin**]: 1 integer, used to reference the Yann LeCun MNIST location of the original digits in the source data files from 0 to 59,999 for train 0 to 9,999 for test or -1 to indicate black signal (meaning a black image not coming from the original MNIST datasets).

[**digit_event**]: 1 integer with the original MNIST label of the image from 0 to 9 or -1 to indicate black signal (no digit shown).

[**original_png**]: 784 integers (comma separated), with the original pixel intensities from the Yann LeCun MNIST from the source png files shown, each pixel can have a value from 0 to 255, (for black signal or event -1 all will be 0s) 784 comes from (28x28) since it is single channel square image, flattened.

[**timestamp**]: 1 Unix Like timestamp for initial time of capture of the signals for this digit event.

[**EEGdata**]:

25,600 floating point (comma separated), sequentially 400 for each of the 64 EEG Channels raw signals (2secs at 200hz) in this order FP1, FPz, FP2, AF3, AFz, AF4, F7, F5, F3, F1, Fz, F2, F4, F6, F8, FT7, FC5, FC3, FC1, FCz, FC2, FC4, FC6, FT8, T7, C5, C3, C1, Cz, CCPz, C2, C4, C6, T8, TP7, CP5, CP3, CP1, CPz, CP2, CP4, CP6, TP8, P7, P5, P3, P1, Pz, P2, P4, P6, P8, PO7, PO5, PO3, POz, PO4, PO6, PO8, CB1, O1, Oz, O2 and CB2.

There are 59,494,400 Data Points, in the main file of this dataset Visual MNIST of Brain Digits using the custom 64 Channels cap.

Additionally, all the signals were postprocessed to generate a Morlet wavelet transform [26] scalogram [27] image for each of the 2,324 rows, using 1 single PNG image per 64 channels related to each digit or black event.

Each color PNG file is 400 pixels width by 6,400 pixels height, using the (x) axis as "time" from left to right and 100 (y) pixels for the resulting Morlet wavelet data for each channel in the above channel order from top of the image to the bottom.

The following image is a sample of just 3 EEG channels of the 64 that are included in each PNG image.

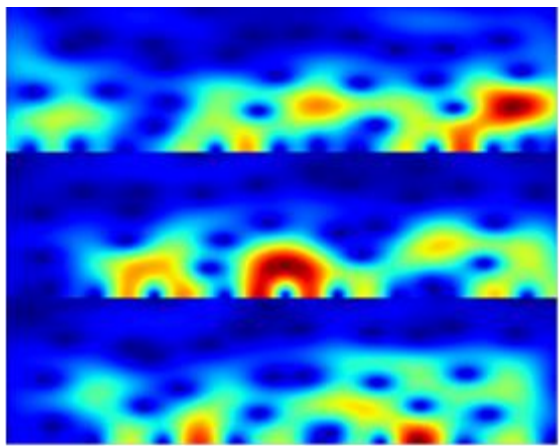

The PNG filenames follow the next naming convention:

A fix initial string "MindBigData64_Mnist2022_"

Followed by "TRAIN_" or "TEST_" referring to the initial Yan LeCun MNIST source of the image shown.

Followed by the [origin] location of the MNIST image with trailing zeros for example "00000_"

And Followed by the [digit_event] 0 to 9 or -1 for example "-1"

Ending with ".png"

For example:

 "MindBigData64_Mnist2022_TRAIN_00000_-1.png"

All the images are inside a folder named for the current version of the dataset "MindBigData64_Mnist2022-Morlet_v0.016"

Note that the -1 events, or black screen, do also include a reference to the MNIST original digit for example "TRAIN_00000" because the black screen brain signal was captured just before the MNIST pixels of this digit was show on the TV screen.

These datasets can be downloaded at:

http://mindbigdata.com/opendb/visualmnist.html

December 2022 curated versions of this datasets at Hugging Face, are explored in the following chapter 2.4.

## 2.4- Hugging Face December 2022 Files

As December 2022, a prepared version of all the previously mentioned datasets was introduced as long with basic Python [27] code to load them.

Also, a selection of 80% train and 20% test distribution was introduced to standardize possible results by the research community using these datasets.

It can be downloaded here:

https://huggingface.co/datasets/DavidVivancos/MindBigData2022

## 3- Constructing MindBigData

The idea behind MindBigData was to somehow replicate the already successful datasets in the Machine Learning and Deep Learning community like the MNIST and ImageNet mentioned above, but using brain activities in the loop, and trying to overcome some of the limitations for capturing huge datasets imposed by the existing brain "reading" technologies.

At the beginning several of these technologies were evaluated, like MRI (like in the picture bellow) that helped determine the brain areas involved in the visualization and thought processes evoked while looking at 0-9 digits.

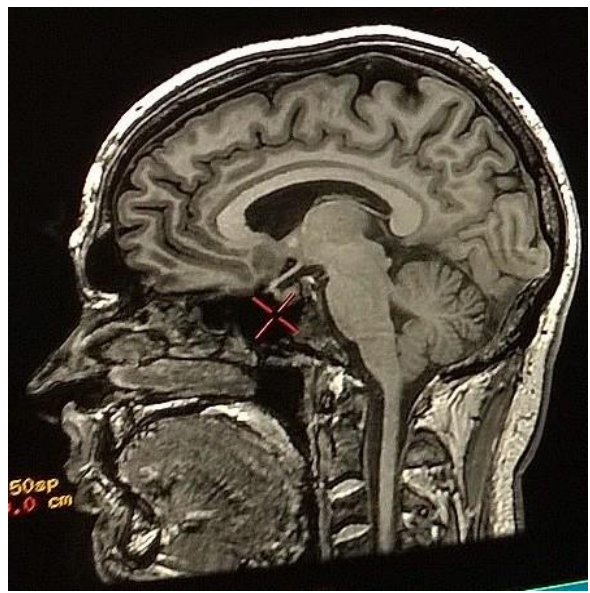

This MRI dataset is not published but it is from the same single subject that "provided" his living brain for the capture of all the EEG signals in the MindBigData datasets, David Vivancos [28], coauthor of this paper. (Male subject born in 1976).

Other technologies were evaluated like FNIRS [29] but were discarded due to readiness of the devices, cost, or complexity to setup in each of the capture sessions.

EEG was selected due the maturity of the technology, being used for more than 100 years, and due to the advances since the beginning of the XXI century, enabling the availability of "commercial off the self" hardware, instead of using expensive medical ones.

Some of these new devices are wireless, enabling brain "reading" capabilities almost everywhere, and most of them use "dry" or "semi-dry" sensors instead of the usual gel based wet ones, used in most medical EEG devices, reducing the setup time from hours or several minutes or seconds in some headsets.

Also, the quality of the signals has been improving overtime, but it is fair to say that noise still can pollute the signals, from muscular to many other body parts movements, electromagnetic interference, or electrodes impedances to name a few, being the signal to noise ratio an issue, and something still to be solved in the field.

Once the technology was selected, custom software was built to capture the signals, and a careful experiment setup was devised.

The setup including a single subject, to reduce possible variations due to brain morphology, gender, or other factors. We know that it is also a limiting factor in terms of the generalization of the results to a wider population, that hopefully could be addressed in future iterations.

Since recordings started in 2014, subject aging could be a factor to consider too, but it was in principle discarded since probably the aging of the devices used to capture could have a more significant impact in the signals, anyway, measuring and possibly correlating this could be something to explore in the future.

Regarding the subject, he was always seated, in the same environment for each recording. The setup was also a fully controlled experiment, carefully trying to avoid, as much as possible, any body movements or blinks. Since these are the main polluting factors in EEG signal recording. The data capture is done in small batches, leaving some time in between to get ready for the next capture.

It is fair to say that some muscular artifacts could still be present in some of the recordings, since even slight muscular tension can alter EEG signals, and the device performance can be degraded over usage time or there can simply be other unaccountable variations in performance, so post processing and filtering should be considered.

Due to the time needed to setup the devices, the subject, the experiment settings and for the physical capture of brain signals, the number of signals to capture was limited, and even so over the course of almost a decade 1,498,089 signals were captured.

Until the beginning of 2022 all the EEG devices used to create MindBigData were commercial wireless devices with great flexibility and mobility capabilities, but they were somehow limited in the number of channels or different areas of the brain covered.

So, in 2022 we begin building our own devices to increase our spatial resolution, starting with the creation of a custom cap with 64 EEG channels, having in mind a setup that could enable the main goals of MindBigData including an easy preparation or time to capture, and readiness of the device that is compatible with a bigger number of capture sessions, since human time is a limiting factor.

Also trying to build a device that could be significantly cheaper than the existing medical devices, achieving this involved deep research in the field, and above all a lot of trial and error until we find the right tools.

It is true that it helped that David Vivancos co-author of this paper has been involved in the Neurotechnology field since 1996 when he created the source code to control a virtual reality world through brain waves, using a single channel wired RS232 [30] EEG, in the picture bellow he is on the left, with Jesus Alido [31] on the right, who is wearing the EEG, renown digital artist and cofounder of his first startups.

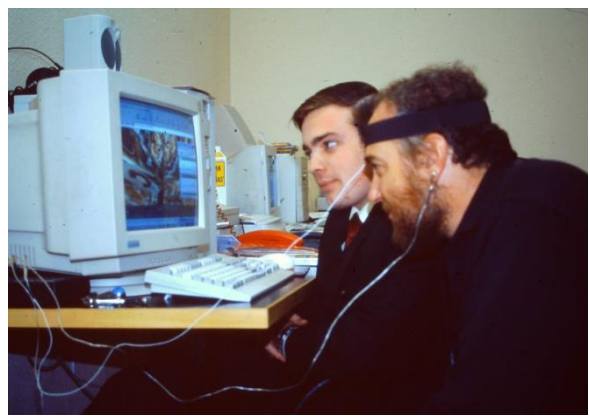

And over the following decades David advised over 10 other initiatives and startups worldview in this space, mostly related to the use of brain signals to create software aimed to tackle specific use cases in several industries and sectors.

Also including the pioneer hardware startup in the field of Brain Computer Interfaces Emotiv Inc. created by Tan Lee [32] and Geoff Mackellar [33], cited in over 15,000 scientific articles.

And since 2022 also helping another promising hardware startup in the field of BCI Nexstem [34] by Siddhant Dangi [35] & Deepansh Goyal [36].

At the end non-invasive technology to read our brains need to evolve to have enough resolution and being wearable enough so the act of recording does not induce discomfort and can be used in as many settings as possible.

But also, the algorithms to translate these signals into the original human intent must also evolve to make the hardware devices usable at a larger range of activities.

And this is the fundamental reason behind building an opening MindBigData, so one day soon we can fully interact with technology evolving from typing to touching to talking to finally using the speed of thought to connect humans to machines.

**4- MindBigData Applications**

Since the beginning of the project almost a decade ago the main driver has been to improve the life of humans, either through helping them to do tasks and actions that they could not do, for any possible reason, or to take to the next level how they do tasks and actions that they already perform increasing accuracy, or simply by making them easier or faster.

Always thinking with an open view, integrating other sciences and disciplines, like robotics, automation, psychology, or healthcare to name a few. As a mean to improve our lives, something we hope subsequent works gets inspiration by this initiative.

We will review a few concrete use cases, not focusing on the datasets per se, but on the possibilities that controlling technology with just our thoughts can bring, since the datasets end goal is enabling the translation from brain signals into repeatable actions.

There are three levels to explore:

1.- Enabling humans to do tasks they can't do with their current capabilities, the first example could be anyone with a handicap, like a tetraplegic, if they could convert their thoughts directly into physical actions, with this they could overcome most of their limitations at personal or professional levels, almost as anyone else without handicap. For this the integration with robotics is needed for example with the help of an exoskeleton, their quality of life could be improved drastically. The World Health Organization estimates that more than 1.3 billion people [37] has some type of handicap in the world.

Another healthcare related use case is improving the lives of millions of children and adults with ADD or ADHD [38].

2.- Improving the accuracy or extending what we currently do, for example again in healthcare, more that 1 million of surgeons [39] could improve the speed and accuracy of its activities, or in many sports where the right mental state has a great impact on performance, or any work in industry or manufacturing were accuracy and speed are paramount, for example we have already witness the use of a robotic arm controlled by Chinese Taikonauts using just their thoughts [40].

3.- By making it easier the tasks that we already do, maybe the most synergistic are the 2.3 billion homes [41], that can benefit in the future by having a brain-controlled automation system, from opening doors, temperature control, control any appliance or rise or lower the blinds to name a few and all controlled by our thoughts.

The last example can be entertainment at large, with gaming with 3 billion users worldwide [42] that could one day play with just their thoughts or interact in any virtual reality environment like in the Metaverse, without the need of any external devices like keyboards, mouses or controllers or even without using their hands, increasing speed and accuracy.

Tu sum up, the objective of this endeavor and following research initiatives to come, is to improve the life of humans, making it easy to interreact mentally with any technology, removing everything in the middle between that now are enablers but also add noise and induce a lost in precision or speed. With the goal of really leveraging the speed of thought.

**5- Discussion and Future Work**

There are several dimensions worth exploring in terms on how to improve the presented datasets:

1.- Increasing the hardware capabilities

2.- As a challenging Benchmark

3.- Increasing the scope

**5.1.- Hardware**

The next step in the evolution of MindBigData is start capturing signals with also a custom-built device but with 128 Channels to expand the spatial coverage of brain activities as much as possible.

The device is already built, and now we are tuning the software side to enable the capture and verifying the signal quality.

So, it is expected that in the following months after the release of this paper new datasets will be included to reflect higher density recordings.

Also new consumer headsets will be used to capture signal, being the next in the pipeline the new Nexstem v2 headset with 19 channels when it is released, probably by the second quarter of 2023.

We are open to include other commercial EEG headsets as long as we can record the raw EEG wave from them.

Another improvement avenue will be to include more capture modalities beyond EEG once the hardware is available at a reasonable cost so it can be mainstream and research at the end can permeate to real products with a benefit for society.

### 5.2.- Benchmark

While the original, non brain related, MNIST or Imagenet datasets are widely used and reached almost peak performance by many talented researchers and labs in the Machine Learning and Deep Learning community over the last decade, kickstarting what some of us call the new golden age of Artificial Intelligence, we cannot say the same for the decoding performance of EEG datasets, using noninvasive recording devices.

The MindBigData datasets are still one of the most challenging open data available for the classification of "brain" digits or "brain" images.

Besides creating a leaderboard with models and performances, it will be desirable to present it or create open competitions with it, and a great place could be any of the main gatherings of the community, like ICML [43] , ICLR [44] , Neurips [45] or CVPR [46].

### 5.3.- Scope

Since atm we have more options to cover a wider range of brain areas, due to higher density devices, will be good to also increase the range of possible datasets beyond the MNIST digits and Images of Imagenet.

So, we are exploring options to include further modalities in the capture pipeline, some still related to pure mental activities maybe involving other senses like audition or some including some simple physical tasks but having in mind that movement induces noise, and the limits of the current devices could be a challenge to overcome.

If the funding will allow it at some point also will enrich the dataset including several more subjects in the dataset, with the challenging task not only of the time needed for the capture, but also ensuring the right setup of the experiment.

### 6.- MindBigData Citations

We wanted to include a list of relevant usages in other scientific articles, publications, or other teaching activities using MindBigdata over the previous years to help continue the research with these technologies and the datasets.

Coming from dozens of universities, institutions, and researchers from more than 16 countries:

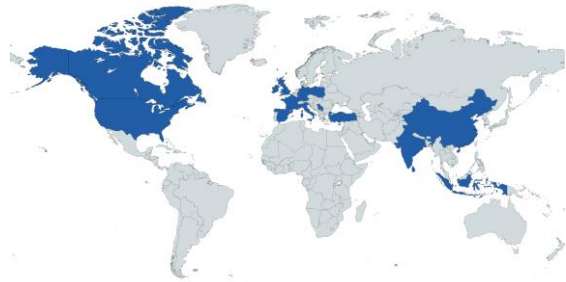

All links last reviewed on December 26th 2022.

### 6.1.- Using the "MNIST" of the Brain Digits Dataset:

- Giving sense to EEG records [47].

- Contributions to fast matrix and tensor decompositions. [48].

https://theses.hal.science/tel-01713104/document

- Fast learning of scale-free networks based on Cholesky factorization [49].

https://onlinelibrary.wiley.com/doi/abs/10.1002/int.21984

- Structured learning from big data based on probabilistic graphical models [50].

https://www.etf.bg.ac.rs/uploads/files/javni_uvid/izvestaji/doktorske/2018/05/Vladisav_Jelisavcic_doktorska_disertacija.pdf

- Combination of Wavelet and MLP Neural Network for Emotion Recognition System [51].

https://www.ijfrcsce.org/index.php/ijfrcsce/article/view/1798

- A Deep Evolutionary Approach to Bioinspired Classifier Optimisation for Brain-Machine Interaction [52].

https://downloads.hindawi.com/journals/complexity/2019/4316548.pdf

- Novel joint algorithm based on EEG in complex scenarios [53].

https://www.tandfonline.com/doi/full/10.1080/24699322.2019.1649078

- HHHFL: Hierarchical Heterogeneous Horizontal Federated Learning for Electroencephalography [54].

https://arxiv.org/pdf/1909.05784.pdf

- Universal EEG Encoder for Learning Diverse Intelligent Tasks [55].

https://arxiv.org/pdf/1911.12152.pdf

- Stanford CS230 - Group Project Final Report, 2020 [56].

http://cs230.stanford.edu/projects_spring_2020/reports/38857160.pdf

- Mental State Recognition and Recommendation of Aids to Stabilize the Mind Using Wearable EEG [57].

https://dl.ucsc.cmb.ac.lk/jspui/bitstream/123456789/4517/1/2017%20MIT%20089.pdf

- Generating the image viewed from EEG signals [58].

https://dergipark.org.tr/tr/download/article-file/1680734

- EEG-Based Emotion Classification for Alzheimer's Disease Patients Using Conventional Machine Learning and Recurrent Neural Network Models [59].

https://www.mdpi.com/1424-8220/20/24/7212

- Understanding Brain Dynamics for Color Perception Using Wearable EEG Headband [60].

https://www.mdpi.com/1424-8220/20/24/7212/pdf

- Frequency Band and PCA Feature Comparison for EEG Signal Classification [61].

https://ojs.unud.ac.id/index.php/lontar/article/view/69866/38951

- Toward lightweight fusion of AI logic and EEG sensors to enable ultra edge-based EEG analytics on IoT devices [62].

https://knowledgecommons.lakeheadu.ca/bitstream/handle/2453/4803/TazrinT2021m-1a.pdf

- Deep Learning in EEG: Advance of the Last Ten-Year Critical Period [63].

https://arxiv.org/pdf/2011.11128.pdf

- Convolutional Neural Network-Based Visually Evoked EEG Classification Model on MindBigData [64].

https://link.springer.com/chapter/10.1007/978-981-16-1543-6_22

- Using Convolutional Neural Networks for EEG analysis [65].

https://github.com/CNN-for-EEG-classification/CNN-EEG

- Visual Brain Decoding for Short Duration EEG Signals [66].

https://ieeexplore.ieee.org/document/9616192

- Quality analysis for reliable complex multiclass neuroscience signal classification via electroencephalography [67].

https://www.emerald.com/insight/content/doi/10.1108/IJQRM-07-2021-0237/full/html

- Toward reliable signals decoding for electroencephalogram: a benchmark study to eegnex [68].

https://arxiv.org/pdf/2207.12369.pdf

**6.2.- Using the "ImageNet" of the Brain Dataset:**

- Inferencia de la Topologia de Grafs [69].

https://upcommons.upc.edu/bitstream/handle/2117/178170/TuraGimenoTFG.pdf

- Understanding Brain Dynamics for Color Perception using Wearable EEG headband [70].

https://arxiv.org/pdf/2008.07092.pdf

- Developing a Data Visualization Tool for the Evaluation Process of a Graphical User Authentication System [71].

https://dms.cs.ucy.ac.cy/op/op.Download.php?documentid=16735&version=1

- Object classification from randomized EEG trials [72].

https://openaccess.thecvf.com/content/CVPR2021/papers/Ahmed_Object_Classification_From_Randomized_EEG_Trials_CVPR_2021_paper.pdf

- Evaluating the ML Models for MindBigData (IMAGENET) of the Brain Signals [73].

http://mindbigdata.com/opendb/2022-ACCS8%20Proceedings%20Book-81-82.pdf

### 6.3.- Extra Literature worth exploring:

EEG Wavelet Classification for Fall Detection with Genetic Programming [74]

Synthetic Biological Signals Machine-generated by GPT-2 improve the Classification of EEG and EMG through Data Augmentation [75]

Decoding Multi-class Motor-related Intentions with User-optimized and Robust BCI System Based on Multimodal Dataset [76]

A large and rich EEG dataset for modeling human visual object recognition [77]

Human EEG recordings for 1,854 concepts presented in rapid serial visual presentation streams [78]

**Acknowledgment**

For any advancement in science and technology to take place we need to have the tools to make it possible, so we need to thank all the brave builders of hardware to read brain activities because they made this possible, and because they keep pushing the limits to increase the resolution, quality, and wearability of the devices, without them MindBigData won't be possible.

Also, thanks to more than 100 researchers in intersection between Machine Learning and Neuroscience that have used MindBigData in their activities because also they contribute to make it better and bring new possibilities.

At the beginning of MindBigData we need to also thank Norberto Malpica [79] and team from the University Rey Juan Carlos [80] for their help with the initial MRI scans.